\documentclass[aps,prd,twocolumn,superscriptaddress,showpacs,showkeys]{revtex4-1}
\usepackage{amsmath}
\usepackage{amsfonts}

\begin{document}
\title{Hawking Radiation as a Probe for the Interior Structure of Regular Black Holes}
\author{Yanbin Deng, Gerald Cleaver}
\affiliation{Department of Physics, Baylor University, Waco,  Texas 76706, USA}
\date{\today}

\begin{abstract}
The notion of the black hole singularity and the proof of the singularity theorem were considered great successes in classical general relativity. Singularities had presented deep puzzles to physicists. Conceptual challenges were set up by the intractability of the singularity. The existence of black hole horizons which cover up the interior, including the singularity of the black hole from outside observers, builds an information curtain, further hindering physicists from understanding the nature of the singularity and the interior structure of black holes. The regular black hole is a concept produced out of multiple attempts of establishing a tractable and understandable interior structure for black hole as well as avoiding the singularity behind the black hole horizon. A method is needed to check the correctness of the new constructions of black holes. After studying the Hawking radiation by fermion tunnelling from one type of regular black hole, structure dependent results were obtained. The result being structure dependent points out the prospects of employing the Hawking radiation as a method to probe into the structure of black holes.
\end{abstract}
\maketitle

\section{Introduction}
The establishment of general relativity(GR) by Einstein was a great success of physics in the last century. GR has become the standard theoretical foundation of the whole modern study of classical gravitational physics from astrophysics to cosmology. Despite great successes, a few difficulties still remain. It is arguable whether we should reserve GR but be foreced to answer the question about the existence and nature of dark matter and dark energy, or to invest on the tireless attempts trying to modify GR to circumvent the concepts of dark matter and dark energy. Nevertheless, the existence of the singularity is an intrinsic problem of GR. The so-called singularity theorem proved by Penrose and Hawking \cite{Penrose&Hawking1, Penrose&Hawking2, Penrose&Hawking3, Penrose&Hawking4} argues that under quite general classical assumptions, the spacetime evolution will inevitably lead to some singularity. It is unfortunate that the geometric curvature blows up and physical laws lose their predictability at the singularities.

At the fundamentally level, the resolution of the singularity problem lies with the expectation that under situations where quantum effects become strong, the behavior of gravity could possibly greatly deviate from that predicted by the classical theory of GR. Various attempts haven been made in exploring the collapse processes and from there seeking for the interior structure of thus produced black holes.\cite{VPFrolov, Kawai&Yokokura} It was James Bardeen who proposed a regular black hole solution which possesses one more parameter than the typical metric of the Reissner-Nordstrom black hole.\cite{Bardeen} This new type of black hole solution shows the same spacetime geometry outside the horizon as the traditional black hole, but bears no singularity inside, therefore endowed the name the {\it regular} black hole.

Bardeen's initiative attracted general intreset and became one of the typical ways of constructing regular black hole solutions. Several similar black hole solutions were proposed\cite{MarsMartin-Prats, CaboAyn-Beato, Borde1, Borde2, BarrabsFrolov, Ayn-BeatoGarcia}. Some of those efforts offer regular black hole solutions, but without specifying the energy-momentum sources. Some are sourced by a nonliear electromagnetic field energy-momentum. The nature of these regular black holes in the region far from the center is the same as traditional black holes, but the property is greatly different near the position where the singularity of traditional black holes locates. The geometric characteristic quantities of those black holes become well-behaved through out the full physical range of parameters. Various additional regular black hole solutions have been constructed along the years by different methods.\cite{Hayward, Linyangzu, Dymnikova1, Dymnikova2, Dymnikova3, Dymnikova4, Dymnikova5, Lilinyang, Linliyang, Lihonglin, Flachilemos}

Whether or not black hole singularities should really exist, they would always be covered up by black hole horizons. Black hole horizons serve as an information curtain hindering outside observers from directly observing the interior structure of the black hole, and determining the eventual existence of the black hole singularities. The so-called black hole no hair theorem served narrowing the choices of methods for pursuing the understanding of the interior structure of black holes. Hawking radiation is one of the rare methods we can apply to infer some but usually little information from the black hole.\cite{Hawking, KrausParikWilczek, SrinivasanPadmanabhan, KernerMann, LinYang, Chen&Huang} But it has not been generally recognized yet that it can be a successful tool for unfolding the interior structure for black holes. We conjecture that in the case of regular black holes, which have rich interior structures to unfold, the Hawking radiation could be a powerful probe into the black hole interior structure.

To test the above conjecture, in this article we make a check and promising results are achieved. To carry out the work, we picked as an example one type of regular black hole with shell-like mass distribution within the horizon. We studied the Hawking radiation through fermion tunnelling from the above type regular black hole, to infer any information about the radiation. The property of the radiation is found to be structure dependent - dependent on the mass distribution. The result being structure dependent encourages the possibility of employing the Hawking radiation as a promising method to probe into the structure of black holes. Our result echoes some attempts in the same direction.\cite{baochengzhang} We present our current result based on a case study and look forward to a general and exact proof in the subsequent research.

\section{One Type of Regular Black Hole}

One type of method of regular black hole solution construction relies on the understanding that there could exist a fundamental minimal length scale arising from some quantum mechanism. Due to the assumption of the existence of a minimal length scale $\theta$ in the noncommutative geometry inspired black hole theory, the mass distribution inside the black hole is not singular, but smeared as\cite{Nicolini1}
\begin{equation}
\rho(r) = \frac{M}{{4\pi \theta}^{3/2}}\exp\left(-\frac{r^2}{4\theta}  \right).\label{smear}
\end{equation}
The Schwarzschild-like black hole in such noncommutative geometry inspired black hole theory  has a metric specified by,
\begin{eqnarray}
ds^2 = &-& \left( 1- \frac{2MG}{\sqrt{\pi}r}\gamma \left(\frac{3}{2},\frac{r^2}{4\theta}  \right) \right)dt^2 \nonumber\\
&+& {\left( 1- \frac{2MG}{\sqrt{\pi}r}\gamma \left(\frac{3}{2},\frac{r^2}{4\theta}  \right) \right)}^{-1}dr^2 + r^2d{\Omega}^2,\label{ds1}
\end{eqnarray}
where $\gamma$ is the so-called lower incomplete Gamma function, with the definition,
\begin{equation}
\gamma \left(\frac{3}{2},\frac{r^2}{4\theta}\right) \equiv \int_{0}^{\frac{r^2}{4\theta}} t^{\frac{1}{2}}e^{-t}dt.
\end{equation}

Many extensions of the above solution eq.\ (\ref{ds1}) are possible. One possible extension is to rotate the black hole up. We summarize such a solution of a Kerr-like black hole with a smeared mass distribution based on the noncommutative black hole theory and we shall use this regular black hole as an example to reveal the structure dependence of Hawking radiation.\cite{Larranaga}

We take the gravitational source to be a smeared mass layer,
\begin{equation}
\rho(r) = Ar^n\exp\left(-\frac{r^2}{l^2}  \right),
\end{equation}
with $l^2=4\theta$ as the convention in P.Nicolini's publication. $A$ is a normalization constant for the mass to be determined by the mass distribution as a function of $r$,
\begin{equation}
m(r) = \int_{0}^{r} 4\pi \bar{r}^2\rho(\bar{r})d\bar{r} = 2\pi A l^{n+3} \gamma \left(\frac{3+n}{2},\frac{r^2}{l^2}\right).
\end{equation}
The Schwarzschild-like black solution is written as,
\begin{equation}
ds^2 = -f(r)dt^2 + \frac{dr^2}{f(r)} +rd\Omega^2,
\end{equation}
where
\begin{equation}
f(r) = 1- \frac{M}{4\pi r \Gamma\left(\frac{n+3}{2} \right)} \gamma \left(\frac{3+n}{2},\frac{r^2}{l^2}\right).
\end{equation}

The Newman-Janis algorithm is then used. This is a method of complex coordinate transformation capable of transforming non-rotating black hole solutions into rotating black hole solutions. A change of coordinates to the outgoing Eddington-Finkelstein coordinates followed by a complex coordinate transformation produces the following Kerr-like black hole with a smeared mass distribution,
\begin{eqnarray}
ds^2 &&= - \frac{\Delta - a^2\sin^2\theta}{\Sigma} dt^2  \nonumber\\
-&& 2 a\sin^2\theta \left( 1- \frac{\Delta - a^2 \sin^2 \theta} {\Sigma} \right) dt d\phi + \frac {\Sigma} {\Delta} dr^2 + \Sigma {d\theta}^2  \nonumber\\
+&& \left[ \Sigma + a^2 \sin^2 \theta \left( 2- \frac {\Delta- a^2 \sin^2\theta} {\Sigma} \right) \right] \sin^2\theta {d\phi}^2,\label{klbh}
\end{eqnarray}
where
\begin{eqnarray}
\Sigma &&= r^2 + a^2 \cos^2\theta,  \nonumber\\
\Delta &&= r^2 - 2 m(r) r + a^2.
\end{eqnarray}

The positions of the horizons of this type of regular black holes are given by $\Delta(r_H)=0$,
\begin{equation}
r_H^2 - \frac{Mr_H}{4\pi \Gamma\left(\frac{n+3}{2} \right)} \gamma \left(\frac{3+n}{2},\frac{r_H^2}{l^2}\right) + a^2 = 0. \label{r_H}
\end{equation}

We see that this equation contains modifications to the traditional equation for the event horizon of a traditional black hole. We will study the Hawking radiation by the fermion tunnelling process for this type of regular black holes and what will turn out in the last is a non-trivial dependence of the Hawking radiation on the modified horizon radius and interior mass distribution function inside the black hole.

\section{Hawking Radiation by Fermion Tunnelling as a Probe for Black Hole Interior Structure}

Before Hawking argued that there should be black body radiation emitted outward from black holes' event horizon, black holes were believed to be completely black. It was Hawking who revealed to people that black holes probably radiate like a black body\cite{Hawking}. In this section we study the Hawking radiation by the fermion tunneling process for the regular black hole whose solution is summarized in section II.

The method we use was proposed, developed, and summarized in\cite{KrausParikWilczek, SrinivasanPadmanabhan, KernerMann, LinYang, Chen&Huang}. The motion of fermions is described by the Dirac equation in curved spacetime,
\begin{eqnarray}
\gamma^\mu D_\mu \psi + \frac{m}{\hbar} \psi = 0, \label{decp}
\end{eqnarray}
\begin{eqnarray}
D_\mu = \partial_\mu + \frac{i}{2} \Gamma^{\alpha \beta}_\mu \Pi_{\alpha \beta} + \frac{iqA_\mu}{\hbar}.
\end{eqnarray}
Here $q$, $m$ and $A_\mu$ are the charge, mass of the fermion and the electric potential in the background respectively. The fermion wave function is given by,
\begin{eqnarray}
\Psi = \psi(t, r,x^\mu)\, e^{\frac{i}{\hbar} S(t, r,x^\mu)}.\label{ps}
\end{eqnarray}
We insert the wave function into the Dirac equation, factor out the exponential terms, and multiply by $\hbar$. Then using the semiclassical approximation (that is, keeping terms only to the leading order in $\hbar$) produces,
\begin{eqnarray}
\gamma^\mu && \left((\partial_\mu \psi)e^{\frac{i}{\hbar}S} + \psi e^{\frac{i}{\hbar} S} \frac{i}{\hbar} \partial_\mu S \right. \nonumber\\
&&\left. + \frac{i}{2} \Gamma^{\alpha \beta}_{\mu} \Pi_{\alpha \beta} \psi e^{\frac{i}{\hbar} S} + \frac{iqA_\mu}{\hbar} \psi e^{\frac{i}{\hbar}S} \right) + \frac{m}{\hbar} \psi e^{\frac{i}{\hbar}S} = 0, \nonumber\\
\gamma^\mu && \left((\partial_\mu \psi) + \psi \frac{i}{\hbar} \partial_\mu S + \frac{i}{2} \Gamma^{\alpha \beta}_{\mu} \Pi_{\alpha \beta} \psi + \frac{iqA_\mu}{\hbar} \psi  \right) + \frac{m}{\hbar} \psi = 0. \nonumber
\end{eqnarray}
This simplifies to,
\begin{equation}
i \gamma^\mu \left( \frac{\partial S}{\partial x^\mu} + q A_\mu \right) \psi + m \mathbf{\psi} = 0.
\end{equation}
Multiplying both sides of this equation by the matrix $ -i \gamma^\nu \left( \frac{\partial S}{\partial x^\nu} + q A_\nu \right)$, and noticing that $ -i \gamma^\nu \left( \frac{\partial S}{\partial x^\nu} + q A_\nu \right)\psi = m \psi$ yields,
\begin{eqnarray}
\gamma^\nu \left( \frac{\partial S}{\partial x^\nu} + q A_\nu \right) \gamma^\mu \left( \frac{\partial S}{\partial x^\mu} + q A_\mu \right)\psi + m^2 \psi = 0.
\end{eqnarray}
Exchanging the index $\mu \longleftrightarrow \nu$, gives
\begin{equation}
\gamma^\mu \left( \frac{\partial S}{\partial x^\mu} + q A_\mu \right)\gamma^\nu \left( \frac{\partial S}{\partial x^\nu} + q A_\nu \right) \psi + m^2 \psi = 0.
\end{equation}
Adding above two equations up, and using the commutative relationship for ferminonic fields $\{ \gamma^\mu, \gamma^\nu \} = 2 g^{\mu \nu}$ then produces,
$$
\{ \gamma^\mu, \gamma^\nu \} \left( \frac{\partial S}{\partial x^\mu} + q A_\mu \right) \left( \frac{\partial S}{\partial x^\nu} + q A_\nu \right) \psi + 2 m^2 \psi = 0.
$$
Or equivalently,
\begin{equation}
2 \Big[ g^{\mu \nu} \left( \frac{\partial S}{\partial x^\mu} + q A_\mu \right) \left( \frac{\partial S}{\partial x^\nu} + q A_\nu \right) +  m^2 \Big]\psi = 0.
\end{equation}
The phase part is separated from the wave function. For the wave function to have a non-trivial solution, we require the vanishing of the phase part of the above equation. This produces the Hamilton-Jacobi equation,
\begin{eqnarray}
 g^{\mu \nu} \left( \frac{\partial S}{\partial x^\mu} + q A_\mu \right) \left( \frac{\partial S}{\partial x^\nu} + q A_\nu \right) +  m^2  = 0.
\end{eqnarray}

We solve this equation in the spacetime of the regular black hole with the smeared mass distribution given in the previous section to study the Hawking radiation of the black hole. We know that this black hole is Kerr-like from outside, but has no singularity inside because of the smeared mass distribution. Reciting the metric of this regular Kerr-like black hole in Boyer-Lindquist coordinates, same as in eq.\ (\ref{klbh})
\begin{eqnarray}
ds^2 && = - \frac{\Delta- a^2\sin^2\theta}{\Sigma} dt^2 \nonumber\\
- && 2 a\sin^2\theta \left(1- \frac{\Delta - a^2\sin^2\theta}{\Sigma} \right) dt d\phi +\frac{\Sigma}{\Delta} dr^2 +  \Sigma d\theta^2 \nonumber\\
+ && \left[\Sigma + a^2 \sin^2 \theta \left(2- \frac{\Delta- a^2\sin^2\theta}{\Sigma} \right)\right]\sin^2\theta d\phi^2, \nonumber
\end{eqnarray}
where
\begin{eqnarray}
\Sigma &=& r^2 + a^2 \cos^2\theta, \nonumber\\
\Delta &=& r^2 - 2 m(r) r + a^2. \nonumber
\end{eqnarray}
Simplification yields,
\begin{eqnarray}
&& ds^2 = - \frac{\Delta- a^2\, sin^2\theta}{\Sigma} dt^2 - 2 a\, sin^2\theta \, \frac{r^2 + a^2 - \Delta}{\Sigma}\, dt\, d\phi \nonumber \\
&& + \frac{\Sigma}{\Delta} dr^2 + \Sigma d\theta^2 + \frac{(r^2 + a^2)^2 - \Delta a^2 sin^2\theta}{\Sigma} sin^2\theta \, d\phi^2.
\end{eqnarray}
In the matrix form, the metric is,
\begin{eqnarray}
g_{\mu \nu} =
\left[
\begin {array}{cccc}
 {-\frac {\Delta -{a}^{2} \sin^{2}\theta}{ \Sigma}}&0&0&-{\frac {a \sin^{2}\theta \left( {r}^{2}+{a}^{2}-\Delta  \right) }{ \Sigma}}
 \\
 \noalign{\medskip}0&{\frac { \Sigma}{\Delta }}&0&0
 \\
 \noalign{\medskip}0&0& \Sigma &0
 \\
 \noalign{\medskip}-{\frac {a \sin^{2}\theta \left( {r}^{2}+{a}^{2}-\Delta  \right) }{ \Sigma}}&0&0&{\frac { ((r^2+a^2)^2-a^2\bigtriangleup\sin^2\theta) \sin^{2}\theta}{ \Sigma}}\end {array} \right].
\nonumber\\
\end{eqnarray}
The inverse metric is then,
\begin{eqnarray}
g^{\mu \nu} = \left[ \begin {array}{cccc}
{-\frac {(r^2 + a^2)^2 - a^2 \Delta \sin^2\theta}{\Delta  \Sigma}}&0&0&{-\frac {a \left( {r}^{2}+{a}^{2}-\Delta  \right) }{\Delta  \Sigma}} \\
\noalign{\medskip}0&{\frac {\Delta }{ \Sigma}}&0&0 \\
\noalign{\medskip}0&0& \frac{1}{\Sigma}&0 \\
\noalign{\medskip}{-\frac {a \left( {r}^{2}+{a}^{2}-\Delta  \right) }{\Delta  \Sigma}}&0&0&{\frac {\Delta - {a}^{2} \sin^{2}\theta}{\Delta  \Sigma \sin^{2}\theta}}
\end {array} \right].
\end{eqnarray}

Now we are ready to expand the Hamilton-Jacobi equation out in the spacetime metric. We also choose a suitable form for the undetermined solution of Hamilton-Jacobi equation. Considering the spherical symmetry of Kerr-like black hole event horizon, separating the variables for the action $S$ as,
\begin{equation}
S = - \omega t + j \phi + R(r) + P(\theta),
\end{equation}
where $\omega$ and $j$ are energy and angular momentum for the particle.

The black hole is uncharged, assuming any electric potential being aroused by any other causes is negligible, $A_\mu=0$. Then the Hamilton-Jacobi equation simplifies,
\begin{eqnarray}
g^{\mu \nu}  \frac{\partial S}{\partial x^\mu}  \frac{\partial S}{\partial x^\nu}  +  m^2  = 0.
\end{eqnarray}
Plugging in the inverse metric, with the undetermined expression for the action $S$, we get the following
\begin{eqnarray}
g^{tt}\, \omega^2 + 2 g^{t \phi}\, \omega j + g^{rr} \left( \frac{\partial R}{\partial r} \right)^2 + g^{\theta \theta} \left( \frac{\partial P}{\partial \theta} \right)^2 \nonumber\\
+ g^{\phi \phi}\, j^2  +  m^2  = 0.
\end{eqnarray}
Completely expanding all the terms yields,
\begin{eqnarray}
&& - \frac{\omega^2 (r^2+a^2)^2}{\Delta \Sigma} + \frac{\omega^2 a^2 \sin^2 \theta}{\Sigma} + 2\omega j a \frac{r^2+a^2}{\Sigma\Delta} \nonumber\\
&& - 2\omega ja \frac{1}{\Sigma} + j^2 \frac{1}{\Sigma \sin^2 \theta} -j^2a^2\frac{1}{\Sigma\Delta} \nonumber\\
&& + \left(\frac{dR}{dr}\right)^2 \frac{\Delta}{\Sigma} +\frac{1}{\Sigma}\left(\frac{dP}{d\theta}\right)^2 +m^2=0.
\end{eqnarray}
We then combine some terms into complete squares,
\begin{eqnarray}
\frac{1}{\Sigma} \left(a\omega\sin\theta - \frac{j}{\sin\theta} \right)^2
- \frac {[{\omega}(r^2+a^2)-ja]^2} {\Sigma\Delta} \nonumber\\
+ \frac{\Delta}{\Sigma}\left(\frac{dR}{dr} \right)^2 + \frac{1}{\Sigma}\left(\frac{dP}{d\theta} \right)^2 + m^2 =0.
\end{eqnarray}
Solving for $\frac{dR}{dr}$, multiplying $\Sigma$, dividing by $\Delta$, results in
\begin{eqnarray}
\left(\frac{dR}{dr}\right)^2 =&& \frac{[{\omega}(r^2+a^2)-ja]^2}{\Delta^2} -\frac{1}{\Delta} \left(a\omega\sin\theta -\frac{j}{\sin\theta} \right)^2 \nonumber\\
&& - \frac{1}{\Delta}\left(\frac{dP}{d\theta} \right)^2 - \frac{m^2\Sigma}{\Delta}.
\end{eqnarray}
We care about the radial part of wave transmission factor $R(r)$ because it determines the radial transmission of the fermions. Only the radially transmitted part of the wave can propagate far enough to reach a distant observer. We will find $R(r)$ by integrating across black hole's outer event horizon $r_H$, thus $\Delta\rightarrow0$. Throwing away lower order terms,
\begin{equation}
\left(\frac{dR}{dr}\right)^2 = \frac{[{\omega}(r^2+a^2)-ja]^2}{\Delta^2}.
\end{equation}
Thus
\begin{equation}
\frac{dR}{dr} =\pm \frac{{\omega}(r^2+a^2)-ja}{\Delta}.
\end{equation}
Defining
\begin{eqnarray}
f(r)=\frac{\Delta}{r^2+a^2},\hspace{2mm}{\rm and}\hspace{2mm} \omega_0=\frac{ja}{{r_H}^2+a^2}
\end{eqnarray}
produces
\begin{equation}
\frac{dR}{dr}= \pm \frac{\omega - \omega_0}{f(r)}.
\end{equation}
We can integrate $R$ along $r$. But Hawking radiation comes from just across the outer event horizon of the black hole, for which $r=r_{+} \pm \epsilon$, where $\epsilon$ is infinitesimal. We care about only the integration coming from $r=r_H$. Then by the residual theorem,
\begin{equation}
R_{\pm}=\pm (\omega - \omega_0)\int_{r_H-\epsilon}^{r_H+\epsilon} \frac{1}{f(r)}dr=\pm i\pi \frac{(\omega - \omega_0)}{f'}.
\end{equation}
$R_+$ and $R_-$ corresponds to incoming and outgoing wave function respectively.

The imaginary part of the action $S$ can be find,
\begin{equation}
ImS=ImR=ImR_+ - ImR_-.
\end{equation}

The tunneling rate for the fermion is:
\begin{eqnarray}
\Gamma && = \exp(-2ImS) = \exp\left(-4\pi \frac{(\omega - \omega_0)}{f'}\right)\nonumber\\
&&=\exp\left(-\frac{(\omega - \omega_0)} {T_H} \right),
\end{eqnarray}
with the surface radiation temperature,
\begin{eqnarray}
T_H=\frac{f'(r_H)}{4\pi}.
\end{eqnarray}
where $\Delta=r^2-2m(r)+a^2$, and $m(r)= 2\pi A\gamma(\frac{n+3}{2}, \frac{r^2}{l^2})$.
Note that eq.\ (\ref{r_H}) is the equation for the position of the horizon of the regular black hole, different from that of the traditional black hole,
\begin{equation}
r_H^2 - \frac{Mr_H}{4\pi \Gamma\left(\frac{n+3}{2} \right)} \gamma \left(\frac{3+n}{2},\frac{r_H^2}{l^2}\right) + a^2 = 0. \ (\ref{r_H}) \nonumber
\end{equation}
We see that the temperature of the Hawking radiation for the regular black hole receives quite complex modifications from that of the traditional black hole,
\begin{equation}
f'(r_H)=\frac{1}{4\pi}\frac{d}{dr}\left(\frac{r^2-2m(r)+a^2}{r^2+a^2} \right)\Big|_{r=r_H}.
\end{equation}

Our analytic analysis thus far offers us meaningful conclusions. Because of the non-singular mass distribution, the Hawking radiation from a Kerr-like regular black hole is modified from that of the traditional Kerr black hole. Further numerical analysis of the fermion tunnelling rate and the Hawking radiation temperature would be rendered to computer software processing. Whether or not there should be richer interior structure for black holes beyond the relatively simpler traditional black holes, the Hawking radiation will prove to be a meaningful method in determining this critical question and vital in telling whether the singularity exactly exists inside the black hole given that the above modifications are testable at astronomical observations.

\section{Summary and Conclusion}

The existence of singularity inside traditional black holes sets up deep conceptual challenges to gravitational physics. Regular black holes are constructions of new types of black holes which reproduce the geometry of traditional black holes outside the event horizon, but offer tractable and understandable interior structure, and avoid the singularity inside the black hole. We present our verification to the conjecture that the Hawking radiation could be a suitable test for the correctness of the regular black hole constructions and a probe into the black hole interior structure.

Our present work provides a positive verification to the above conjecture. Our result shows a non-trivial dependence of the Hawking radiation on the modified horizon radius and interior mass distribution function inside the black hole and confirms the possibility of employing the Hawking radiation as a promising method to probe into the structure of black holes as far as these modifications are testable at astronomical observations.

\begin{acknowledgments}
Authors are grateful to Prof. Anzhong Wang and Drs. V.H. Satheeshkumar, Xinwen Wang, Miao Tian and Chikun Ding for useful discussions.
\end{acknowledgments}

Author email addresses:\\
yanbin\_deng@baylor.edu\\
gerald\_cleaver@baylor.edu

\end{document}